\documentclass[preprint,aps, nofootinbib]{revtex4}

\def\be{\begin{equation}}
\def\ee{\end{equation}}
\def\ba{\begin{eqnarray}}
\def\ea{\end{eqnarray}}

\def\dalemb#1#2{{\vbox{\hrule height.#2pt
        \hbox{\vrule width.#2pt height#1pt \kern#1pt \vrule width.#2pt}
        \hrule height.#2pt}}}

\def\gtorder{\mathrel{\raise.3ex\hbox{$>$}\mkern-14mu
             \lower0.6ex\hbox{$\sim$}}}
\def\ltorder{\mathrel{\raise.3ex\hbox{$<$}\mkern-14mu
             \lower0.6ex\hbox{$\sim$}}}

\usepackage[dvips]{graphics,epsfig}
\usepackage{amsmath}
\usepackage{amsthm}
\usepackage{amssymb}
\usepackage{amsfonts}
\usepackage{mathrsfs}

\def\dim#1 {\,^{(#1)}}
\def\versor#1 {\vec{e}_#1}
\def\grad#1 {\nabla_{\vec{e}_#1}}

\def\mpf {M_{P(5)}}

\def\ere {\mathbb R^4}
\def\aalpha {\mpf^3}

\def\be{\beta}

\def\frac#1#2{{\textstyle{{#1}\over {#2}}}}

\def\lsim{\mathrel{\rlap{\lower4pt\hbox{\hskip1pt$\sim$}}
    \raise1pt\hbox{$<$}}}
\def\gsim{\mathrel{\rlap{\lower4pt\hbox{\hskip1pt$\sim$}}
    \raise1pt\hbox{$>$}}}
\def\sqr#1#2{{\vcenter{\vbox{\hrule height.#2pt
         \hbox{\vrule width.#2pt height#1pt \kern#1pt
         \vrule width.#2pt}
         \hrule height.#2pt}}}}
\def\square{\mathchoice\sqr66\sqr66\sqr{2.1}3\sqr{1.5}3}
\begin{document}

\rightline{DF/IST-3.2008}
\rightline{LPT/08-43}

\title{
A new source for a brane cosmological constant from a modified
  gravity model in the bulk
}

\author{Orfeu Bertolami${}^{1,2}$,
Carla Carvalho${}^{1,2,3}$
and Jo\~ao N. Laia${}^{1}$\\
${}^{1}${\it Departamento de F\'\i sica, Instituto Superior T\'ecnico \\
Avenida Rovisco Pais 1, 1049-001 Lisboa, Portugal} \thanks{DF/IST-X.2008}\\
${}^{2}${\it Centro de F\'{i}sica dos Plasmas, Instituto Superior T\'ecnico \\
Avenida Rovisco Pais 1, 1049-001 Lisboa, Portugal} \\
${}^{3}${\it Laboratoire de Physique Th\'eorique, Universit\'e Paris-Sud, \\
B\^atiment 210, 91405 Orsay, France} \thanks{LPT/08-43}\\
%\email{orfeu@cosmos.ist.utl.pt}, \, \email{ccarvalho@ist.edu}, \,
%\email{jnlaia@gmail.com}
}

\begin{abstract}%
{We show that a four-dimensional equation of state for a
cosmological constant term arises from a perfect fluid in the bulk  in
the context of a gravity model where the scalar curvature is
non-minimally coupled to the perfect fluid Lagrangian density.
The four-dimensional theory is fully determined from the induced
equations on the brane, subject to the boundary conditions derived
across the brane.}
\end{abstract}

%\keywords{Braneworld Gravity, Modified Theories of Gravity, Cosmological Term}

%\begin{document}
\maketitle

\vfill

\eject

\section{Introduction}

Braneworld scenarios are an interesting development in what concerns
gravity models and their cosmological implications
\cite{Maartens:2003tw}. Most often these scenarios assume, given
observational constraints as well as theoretical assumptions, that
the bulk space is empty except for a cosmological constant.
However, more recently the implications of having vector and
scalar fields in the bulk were studied in connection with
Lorentz symmetry \cite{Bertolami:2006bf} and gauge symmetry
breaking \cite{Bertolami:2007dt}.

In this paper, we consider the presence of a perfect fluid in the bulk in the
context of a five-dimensional braneword
model where the scalar curvature couples non-minimally to
the Lagrangian density of the perfect fluid.
In (3+1) dimensions this class of models with Lagrangian density of the
form \cite{Bertolami:2007gv}
\ba
\mathcal L = \alpha f_1 (R)
+\left( 1 + \lambda f_2 (R) \right)\mathcal L_M ~ ,
\ea
where $f_1 (R)$ and $f_2 (R)$ are generic functions of the scalar
curvature, %. This novel gravity model
was shown to exhibit interesting features which allow one to
address problems such as the rotation curves of galaxies
without the need of dark matter
(see Ref.~ \cite{Bertolami:2007gv} and references therein)
and the Pioneer anomaly 
(see Ref.~ \cite{Anderson:2001sg, Bertolami:2003ui} and references
therein). The stability of these models has been examined 
in Ref.~\cite{Faraoni2007}. Other studies on their implications
included their impact on stellar equilibrium and the analysis of their
corresponding PPN parameters, which were studied in
Refs.~\cite{Bertolami-Paramos2008a,Bertolami-Paramos2008b},
respectively. 

Recently there has also been interest in the conformal equivalence
between $f(R)$ theories and Einstein gravity non-minimally
coupled to a scalar field in the context of braneworlds
\cite{Deruelle07}.
The expected increasingly higher order of the discontinuity of the
geometric quantities across the brane with the increasing power in $R$
of $f(R)$ is solved by enforcing continuity of the metric to
correspondingly higher-order derivatives.
Here, however, we will not impose further continuity conditions on the
intervening fields, allowing for the discontinuity of the second
derivative of the metric across the brane and orthogonal to its
surface, despite also obtaining an increase in the power of $R.$

Crucial in the setting of our problem is a suitable implementation of
the Israel matching conditions in the presence of bulk fields in order
to extract the boundary conditions, 
both for gravity and the matter fields, which the
induced equations of motion on the brane must satisfy.
The method to be employed here was first introduced in
Ref.~\cite{Bucher:2004we} and further developed in
Refs.~\cite{Bertolami:2006bf,Bertolami:2007dt}.
For completeness and clarity, the more involved
technical details of our method are presented in the Appendix. As we
shall see, and rather remarkably, the projection of the bulk perfect fluid
induces on the brane a new cosmological constant term. This new source
for a brane cosmological constant opens quite interesting
perspectives for inflation at the early universe and for acceleration
at the late time expansion of the universe. This result suggests that
a perfect fluid in the bulk space may have a bearing on the
cosmological constant problem on the brane.

This paper is organized as follows. In section \ref{sec:model} we
present our model and work out a suitable Lagrangian density for a
perfect fluid. This development extends the approach of Hawking and
Ellis \cite{Hawking:1973uf} to the bulk space.
In section \ref{sec:matchingconditions}, we work out
the matching conditions across the brane and derive the equations of
motion therein induced.
A derivation of the Gauss-Codacci relations is also presented in the
Appendix for completion.
Section \ref{sec:results} contains our results and section
\ref{sec:conclusions} our conclusions.

\section{A Modified Gravity Model in the Bulk
\label{sec:model}}

\subsection{The Einstein Equations}
We consider the particular case of the action discussed in
Ref. \cite{Bertolami:2007gv}. We set $f_1 (R) = f_2 (R) = R$ and
introduce a cosmological constant as follows
\ba
S = \int d^5 x \sqrt{-g}
 \left[\aalpha R +(1 +\lambda R )\mathcal L_M  -2\mathcal L_\Lambda
 \right] ~.
\ea
Here $M_{P(5)}$ is the five-dimensional Planck mass, ${\cal L}_{M}$
and ${\cal L}_{\Lambda}$ are respectively the matter and the
cosmological constant Lagrangian densities.

We define the stress-energy tensor as usual
\ba
T_{\mu\nu} = -{2\over\sqrt{-g}}
 {\delta\left(\sqrt{-g}\mathcal L_M \right)\over\delta g^{\mu\nu}} ~
\label{eq:Tmunu}.
\ea
For convenience, we define also the vacuum energy tensor as
\ba
\Lambda_{\mu\nu} = -{2\over\sqrt{-g}}
 {\delta\left(\sqrt{-g}\mathcal L_\Lambda \right)\over\delta g^{\mu\nu}} ~,
\ea
assumed to be of the form
\ba
\Lambda_{\mu\nu} =\Lambda_{(5)} g_{\mu\nu}
+\sigma\delta(N)( g_{\mu\nu} -N_\nu N_\mu )~,
\ea
so as to include both the bulk vacuum value $\Lambda_{(5)}$ and the
brane tension $\sigma.$
Here $N_\mu$ are the components of the unit five-vector orthogonal to the
brane $\versor N = N^\mu \versor \mu.$~
Thus, in Gaussian coordinates, the
cosmological constant tensor is given by:
\ba
\mathbf\Lambda =  \left(\begin{array}{ccc|c}
 & & & \\
 & \mathbf{\dim 4 g} \left( \Lambda_{(5)} + \sigma\right)& & \\
& & & \\\hline
& & & \Lambda_{(5)}
\end{array}\right) ~,
\ea
The five-dimensional Einstein equation is obtained by varying the
action with respect to the metric, finding that
\ba
\aalpha G_{\mu\nu} 
-{1\over 2}\left( 1 +\lambda R\right) T_{\mu\nu}
+\Lambda_{\mu\nu}
-\lambda \left( 
 \nabla_\mu \nabla_\nu -g_{\mu\nu} \square 
 -\lambda R_{\mu\nu} 
\right)\mathcal L_M =0 ~.
\label{eq:Einstein5dim}
\ea

\subsection{A Perfect Fluid in the Bulk}

Since the Einstein equation in Eq.~(\ref{eq:Einstein5dim})
contains terms in $\mathcal L_M,$ we must construct a Lagrangian density
associated with a perfect fluid so that by Eq.~(\ref{eq:Tmunu}) it
yields
\begin{equation}
 T_{\mu\nu} = (\rho + p)u_\mu u_\nu + p g_{\mu\nu} ~,
\end{equation}
where $\rho$ is the energy density, $p$ the pressure and $u_\mu $ the
unit five-velocity of the fluid (tangent to the flow lines and thus
time-like, $u_\mu u^\mu = -1$).
For this purpose, we follow closely the procedure described
%by Hawking and Ellis
in Ref.~\cite{Hawking:1973uf}. 

Let the perfect fluid Lagrangian density $\mathcal L_M$ be given by
\ba
\mathcal L_M = -\tau (1+\epsilon ) ~,
\ea
where $\tau$ is an auxiliary variable and $\epsilon$ is the internal
energy of the fluid as well as a function of $\tau.$ Assuming that
the fluid current vector $j^\mu = \tau u^\mu$ is conserved,
$\nabla_\mu j^\mu = 0$,
then $\delta (\sqrt{-g}j^\mu ) = 0$ when the metric is varied.
Then, from
\ba
\tau^2 =-j^\mu j^\nu g_{\mu\nu}
={1\over g}
 \left( \sqrt{-g}j^\mu\right)
  \left( \sqrt{-g}j^\nu\right)g_{\mu\nu}~,
\ea
it follows that
\ba
2 \tau \delta \tau
&=&-{\delta g\over g^2}\left( \sqrt{-g}j^\mu\right)
 \left( \sqrt{-g}j^\nu\right)g_{\mu\nu}\cr
&&+{1\over g}\left( \sqrt{-g}j^\mu\right)\left( \sqrt{-g}j^\nu\right)
 \delta g_{\mu\nu}\cr
&=&\left( j_\mu j_\nu -j^\beta j_\beta g_{\mu\nu}\right)\delta g^{\mu\nu} ~,
\ea
and consequently that the variation of $\tau$ with respect to the
metric is given by
\ba
\delta \tau
={1\over 2}\left( \tau g_{\mu\nu} +\tau u_\mu u_\nu\right)
  \delta g^{\mu\nu} ~.
\ea
Using the definition of the stress-energy tensor,
Eq.~(\ref{eq:Tmunu}), we find that
\ba
T_{\mu\nu}
&=&\mathcal L_M g_{\mu\nu}
-2{\delta \mathcal L_M \over \delta g^{\mu\nu}}\cr
&=&\left[ \tau( 1 +\epsilon) +\tau^2{d \epsilon\over d \tau}\right]
 u_\mu u_\nu
+\tau^2 {d \epsilon\over d \tau}g_{\mu\nu} \cr
&=& (\rho + p)u_\mu u_\nu + p g_{\mu\nu} ~,
\ea
where we made the following identifications
\ba
\rho =\tau (1+\epsilon)~, \qquad
p =\tau^2 {d \epsilon\over d \tau} \label{eq:ident2} ~.
\label{eq:rho+p}
\ea
We have thus obtained the stress-energy tensor for a perfect fluid
from the Lagrangian density $\mathcal L_M = - \rho$ and from the
continuity equation $\nabla_{\mu}(\tau u^{\mu})=0.$
For an alternative formulation where
the Lagrangian density is identified with the pressure see, for
instance, Refs. \cite{Schutz1970,Brown1993}. 

%\subsubsection{Equation of Motion}
To obtain the equation of motion for the perfect fluid, one could
compute the divergence of the Einstein equation, Eq.~(\ref{eq:Einstein5dim}).
%\cite{Bertolami:2007gv}.
Alternatively, we will proceed directly from the Lagrangian
density. For this purpose, we consider the action of the Lie
derivatives on the fluid flow lines.
Let $\gamma :[a,b]\times \mathcal N \rightarrow \mathcal D \subset
\mathcal M$ be a congruence of flow lines, one through each point of
$\mathcal M,$ where $[a,b]$ is the interval of the parameter ascribed
to the flow lines, $\mathcal N$ is some four-dimensional manifold
and $\mathcal D$ is a small region of the five-dimensional spacetime
manifold $\mathcal M$. The tangent vector to the flow lines is given by
$\textbf U =(\partial / \partial t)_\gamma,$ with $t \in [a,b],$ which 
once normalized we identify with the fluid velocity
\ba
u^\mu ={U^{\mu}\over \sqrt{-g_{\alpha\beta}U^\alpha U^\beta}}
={U^\mu\over |U|} ~.
\ea
The action $S$ is required to be stationary when the flow lines are
varied. Variations
with respect to the flow lines, which we here represent by $\Delta,$
amount to variations along the corresponding tangent vectors.
Considering
$\gamma(r,[a,b],\ere)$, where $r$ is the parameter that selects
different congruences of flow lines, then
$\Delta \textbf U =\mathcal L_{\textbf V} \textbf U$,
where $\textbf V = (\partial /\partial r)_\gamma$
and $\mathcal L_{\textbf V}$ is the Lie derivative along $\textbf V$.
Since
$\Delta ( u^\mu |U|) =\mathcal L_{\textbf V} U,$
then
\ba
\Delta u^\mu ={1\over |U|}\left( \mathcal L_{\textbf V} U^\mu
-u^\mu\Delta |U| \right) ~.
\ea
Moreover,
\ba
\Delta |U| =-{1\over|U|}g_{\alpha\beta} U^\alpha \Delta U^\beta
=-g_{\alpha\beta} u^\alpha \mathcal L_{\textbf V} U^\beta
\ea
and
\ba
\mathcal L_{\textbf V} U^\beta
=V^\sigma {U^\mu}_{;\sigma} - U^\sigma {V^\mu}_{;\sigma} ~,
\ea
and consequently
\ba
\Delta u^\mu
=V^\sigma {u^\mu}_{;\sigma} -u^\sigma {V^\mu}_{;\sigma}
-u^\mu u^\beta V_{\beta ;\sigma} u^\sigma ~.
\label{eq:deltau}
\ea
From the conservation of the fluid current vector ${j^\mu}_{;\mu}= 0,$
it follows that $\Delta ({j^\mu}_{;\mu}) =0 =(\Delta j^\mu)_{;\mu}$
which, using Eq.~(\ref{eq:deltau}) and integrating
along the flow lines, yields
\ba
\Delta \tau =(\tau V^\beta)_{;\beta} +\tau V_{\beta;\alpha}u^\beta u^\alpha ~.
\ea
Thus, in the Lagrangian density, $\tau$ varies so that the associated
current vector $j^\mu$ is conserved.

Finally, the condition for the stationarity of the metric yields
\ba
{\partial S\over \partial \tau}
&=&\int \sqrt{-g}d^5x \left\{ {d\mathcal L\over d\tau}\Delta \tau\right\}\cr
&=&\int \sqrt{-g}d^5x \left\{ -(1 +\lambda R)\left[
 1 +{d(\tau \epsilon)\over d\tau} \right] \Delta \tau \right\}\cr
&=&\int \sqrt{-g}d^5x \left\{ -(1+\lambda R)\left[
 1 +{d(\tau \epsilon)\over d\tau} \right]
  \left[ \nabla_\mu  \left( \tau V^\mu \right)
  +\tau \left( \nabla_\beta V_ \mu \right) u^\mu u^\beta \right] \right\}=0 ~.
\ea
Integrating by parts and using the Stokes theorem to discard the
surface terms, we find that
\ba
&&\int \sqrt{-g}d^5x  \left\{ \tau \nabla_\mu \left[
 (1 +\lambda R)\left( 1 +{d(\tau \epsilon)\over d\tau}\right)\right]\right.\cr
&& \qquad  + \left. \nabla_\beta \left[
 (1 +\lambda R)\left( 1 +{d(\tau \epsilon)\over d\tau} \right)\tau
  u_\mu u^\beta\right]\right\} V^\mu =0 ~,
\ea
which holds for any vector $\textbf V.$ Therefore, the expression
within curly brackets must vanish
\ba
&&\left\{ \lambda \nabla_\beta R
   \left[ 1 +{d(\tau \epsilon)\over d\tau}\right]
+( 1 +\lambda R)
  \left[ {d(\tau \epsilon)\over d\tau}\right]_{;\beta}\right\}
\left( g^{\beta\mu} + u^\beta u^\mu\right) \tau \\
&+&\left( 1 +\lambda R\right)
 \left[
  1 +{d(\tau \epsilon)\over d\tau}\right]u^\beta (\nabla_\beta u^\mu)\tau =0~.
\ea
Using the equations in Eq.~(\ref{eq:rho+p}) and noticing that
\ba
\tau \left[ {d(\tau\epsilon)\over d\tau}\right]_{;\beta}
=\left[\tau^2 {d\epsilon\over d\tau} \right]_{;\beta}
=\nabla_\beta p ~,
\ea
we obtain the equation of motion for the perfect fluid in the bulk
\ba
\left[ \lambda\left( \rho +p\right)\nabla_\beta R
 +( 1 +\lambda R)\nabla_\beta p\right]( g^{\beta\mu} +u^\beta u^\mu )
+( 1 +\lambda R)( \rho +p)u^\beta \nabla_\beta u^\mu =0~.
\label{eq:fluid}
\ea
For $\rho +p \neq 0,$  this equation can be rewritten as
\ba
u^\beta \nabla_\beta u^\mu ={D u^\mu\over ds}
={du^\mu\over ds} +\Gamma^{\mu}_{\alpha\beta}u^\alpha u^\beta =f^\mu~,
\ea
where $f^\mu$ can be regarded as an exterior force given by
\ba
f^{\mu} =-\left[ {\lambda\over 1 +\lambda R}\nabla_\beta R
+{1\over \rho +p}\nabla_\beta p\right]\left( g^{\beta\mu}
+u^\beta u^\mu\right)~.
\label{eq:eqmotion}
\ea
Setting $\lambda = 0,$ one recovers the known equation of motion for a
perfect fluid in General Relativity. Moreover, Eq.~(\ref{eq:eqmotion})
agrees with the result obtained in Ref.~\cite{Bertolami:2007gv} by
taking the divergence of the Einstein's field equation. This indicates
that the assumed conservation of the fluid current vector
$j^\mu =\tau u^\mu$ is a consistent description of our physical system.

The continuity equation, which follows from the conservation of $j^{\mu},$
provides the last equation and ensures that our problem
is well defined. From Eq.~(\ref{eq:rho+p}) we find that
\ba
p =\tau^2 \epsilon_{,\mu} {1\over \tau_{,\mu}}, \qquad
\rho {\epsilon_{,\mu}\over 1 +\epsilon}( p +\rho) =\rho_{,\mu}p ~.
\ea
The continuity equation
\ba
{ j^\mu}_{;\mu} =\tau_{,\mu}u^\mu +\tau {u^\mu}_{;\mu}
\ea
is thus equivalent to
\ba
\rho_{,\mu} u^\mu + (p+\rho){u^\mu}_{; \mu} = 0 ~.
\label{eq:continuity}
\ea
For the ideal fluid, the gravitational field enters only in
Eq.~(\ref{eq:fluid}). Gravity does not appear in the velocity
equation, Eq.~(\ref{eq:continuity}), as the velocity is measured
relative to freely moving observers \cite{Peebles80}.

\section{The Induced Equations on the Brane
\label{sec:matchingconditions}}

In this section, we derive the equations of motion induced on the brane.
First, we rewrite the components of the equations of motion derived in
the previous section in Gaussian normal coordinates. 
In our notation, the directions
parallel to the brane are denoted by $A, B,\dots$, while the normal
direction is denoted by $N$ so that the brane is localized at
$N=0.$
Using the results derived in Appendix \ref{sec:decomposition}, we
obtain the relevant components of the Einstein equation:
\ba
&&\aalpha \dim 5 G_{AB}
-{1\over 2}\left( 1 +\lambda \dim 5 R\right)
  \left[ (\rho +p)u_A u_B +pg_{AB}\right]
+\Lambda_{AB} \cr
&+&\lambda \left[ \nabla_A\nabla_B +K_{AB}\nabla_N 
-g_{AB}\left( \square +K\nabla_N +\nabla_N\nabla_N \right)
-\dim 5 R_{AB}
\right]\rho = 0 ~,
\label{eq:einstein:ab}
\\
\cr
%\ea
%\ba
&&\aalpha \dim 5 G_{AN} 
-{1\over 2}\left( 1 +\lambda \dim 5 R\right)
 \left[ (\rho +p)u_A u_N\right] \cr
&+&\lambda \left[ \nabla_A \nabla_N -K_{A}^B\nabla_B 
-\dim 5 R_{AN}\right] \rho=0 ~,
\label{eq:einstein:an}
\\
\cr
%\ea
%\ba
&&\aalpha \dim 5 G_{NN}
-{1\over 2}\left( 1 +\lambda \dim 5 R\right)
  \left[ (\rho +p)u_N^2 +p g_{NN}\right]
+\Lambda_{(5)} \cr
&+&\lambda \left[ \nabla_N \nabla_N \rho
-g_{NN}\left( \square  +K\nabla_N  +\nabla_N\nabla_N \right)
-\dim 5 R_{NN}\right]\rho = 0 ~.
\label{eq:einstein:nn}
\ea
%Note that only on five-dimensional terms do we keep the index
%indicating the corresponding dimension and only whenever confusion is
%propitiated.
Note that 
we will only keep the index indicating the corresponding dimension
on the five-dimensional terms and whenever confusion is propitiated.

We then proceed to derive the matching conditions, which follow from the
presence of the brane dividing into two regions the bulk spacetime and
from the symmetries of the bulk fields across the two regions.
Here we regard the brane as a $\mathbb{Z}_{2}$--symmetric surface of
infinitesimal thickness $2\delta$ and thus
separating the bulk into two mirroring regions about $N=0.$
%In order to guarantee the continuity of quantities on the brane,
The symmetry about the brane establishes how bulk quantities relate on
the two sides of the brane.
Hence, vector components parallel to the brane are even in $N,$
whereas normal components are odd. For tensor quantities this
generalizes by considering each additional normal component to reverse
the parity of the component with one less normal component.
Accordingly, it follows that $u_{A}(N=-\delta) =u_{A}(N=+\delta),$
whereas $u_{N}(N=-\delta)=-u_{N}(N=+\delta).$
Likewise, $g_{AB}(N=-\delta)=g_{AB}(N=+\delta)$ and
$K_{AB}(N=-\delta)=-K_{AB}(N=+\delta).$
Consequently, there will be quantities that are discontinuous across
the brane and whose derivatives in $N$ generate singular distributions
at the position of the brane. Integration of these contributions in
the coordinate normal to the brane allows to relate the induced
geometry of the brane with the induced stress-energy therein
localized. Hence, by extracting the singular contributions from the projected
bulk equations and establishing the matching conditions, we obtain the
equations of motion induced on the brane.

%[the extension of Gauss theorem to forms applied to covariant
% derivatives]

We equate the $(AB)$ component of the Einstein equation
using the Gauss-Codacci conditions in
Eqs.~(\ref{eq:gc1}--\ref{eq:gc3}) as well as Eq.~(\ref{eq:rab}). 
Integrating along the $N$ direction across the
position of the brane, we find that
\ba
-\sigma g_{AB}
&=&\lim_{\delta \rightarrow 0} \int^{+\delta}_{-\delta} dN
\nabla_{N}\biggl\{
\aalpha \left( -K_{AB} +g_{AB}K\right)\cr
&&+\lambda K\left[ ( \rho +p)u_{A}u_{B} +pg_{AB}\right]
-\lambda \left( g_{AB}\nabla_{N} -K_{AB}\right)\rho
\biggl\},
\ea
where upon integration by parts
non-singular terms arise which vanish upon integration and which contribute
to the effective equation of motion. For that we assume
the energy density, the pressure and the fluid velocity to be
continuous. This implies that only
higher than second order derivatives in the $N$ direction can be singular
on the brane and consequently survive over the infinitesimal
integration. From these considerations there follows the Israel
matching condition 
\ba
\left(\aalpha -\lambda \rho\right)\left( -K_{AB} +g_{AB}K\right)
+\lambda K(\rho +p)\left( u_{A}u_{B} +g_{AB}\right)
-\lambda g_{AB}\nabla_{N}\rho
=-{1\over 2}\sigma g_{AB}
\label{eq:israel:ab}
\ea
which, upon taking the trace, yields
%Moreover, we observe that the trace of Eq.~(\ref{eqn:srael:ab})
%provides the relation between $K =g_{AB}K^{AB}$ and $\nabla_{N}\rho.$
%The trace of the extrinsic curvature
\ba
\left(\aalpha -\lambda \rho\right)K( -1+d)
+\lambda K( \rho +p)\left( u_{C}^2 +d\right)
-\lambda d\nabla_{N}\rho
=-{d\over 2}\sigma~.
\label{eqn:K}
\ea
Another useful result is
\ba
\left(\aalpha -\lambda \rho\right)K_{AB}
=K{1\over d}\left[
g_{AB}\left(\aalpha -\lambda \rho\right)
+\lambda(\rho +p)\left( d u_{A}u_{B} -g_{AB}u_{C}^2\right)\right].
\label{eq:useful}
\ea
Substituting Eq.~(\ref{eq:israel:ab}) back in the $(AB)$ component of the
Einstein equation, we obtain
\ba
&&
\left( M_{P(5)}^{3} -\lambda \rho\right)
 \left[ G_{AB} -KK_{AB}
 +{1\over 2}g_{AB}\left( K^2 +K_{CD}K^{CD}\right)\right]
+g_{AB}\Lambda_{(5)}
\cr
&-&\left[
{1\over 2}\left(1 +\lambda\left\{ R -K^2 -K_{CD}K^{CD}\right\}\right)
+\lambda K\nabla_{N}\right]
 \left[ ( \rho +p)\left( u_{A}u_{B} +g_{AB}\right)\right]\cr
&+&\left[
{1\over 2}g_{AB}
+\lambda\left( \nabla_{A}\nabla_{B} -g_{AB}\nabla_{C}^2\right)\right]\rho
=0~.
\label{eq:ab}
\ea

From the $(AN)$ component of the tensor equation we notice that
\ba
G_{AN}= K_{AB|B} -K_{|A}= -\nabla_{B}\int dN G_{AB}
=-\nabla_{A}\cal T_{AB}~,
\label{eq:an}
\ea
where $\cal T_{AB}$ stands for the stress-energy tensor induced on the
brane as given by the Israel matching condition in
Eq.~(\ref{eq:israel:ab}).
If we impose conservation of energy on the brane, it follows that
$G_{AN}=0,$ which implies the condition
\ba
\nabla_{A}\left[ K(\rho +p)(u_{A}u_{B} +g_{AB})\right]
-\left[
g_{AB}\nabla_{N}
-\left( K_{AB} -g_{AB}K\right)\right]\nabla_{A}\rho
=0~.
\label{eq:conservation}
\ea

Furthermore, equating the $(NN)$ component of the Einstein equation
%the Gauss-Codacci relations Eqs.~(\ref{eq:gc1}--\ref{eq:gc3}).
and integrating along the $N$ direction, we find that
\ba
0 &=&\lim_{\delta \rightarrow 0} \int^{+\delta}_{-\delta} dN
\nabla_{N}\left\{
\rho K +K\left[ (\rho +p) u_N^2+p \right] \right\}
 =-2K (\rho + p)u_{C}^2 ~.
\label{eq:israel:nn}
\ea
Substituting back in the $(NN)$ component of the Einstein equation, we
find that
\ba
&&\left( M_{P(5)}^{3} -\lambda \rho\right)
 {1\over 2}\left( -R +K^2 -K_{CD}K^{CD}\right)
+\Lambda_{(5)}\cr
&-&\left[
{1\over 2}\left(1 +\lambda\left\{ R -K^2 -K_{CD}K^{CD}\right\}\right)
+\lambda K\nabla_{N}\right]
 \left[ ( \rho +p)\left( u_{N}^2 +1\right)\right]\cr
&+&\left[
{1\over 2}
-\lambda\left( \nabla_{C}^2 +K\nabla_{N}\right)\right]\rho
=0~.
\ea
Moreover,
substituting $\nabla_{N}\rho$ from the Israel matching condition
for a time-like fluid velocity normalized so that
$u_{A}^2+u_{N}^2=-1,$ it follows that
\ba
&&\left( M_{P(5)}^3 -\lambda \rho\right)
 {1\over 2} 
  \left[ -R +K^2\left( {2\over d} -1\right)-K_{CD}K^{CD}\right]
+\Lambda _{(5)}\cr
&+&\left[ {1\over 2}\left(1 +\lambda \left\{R -K^2 -K_{CD}K^{CD}\right\}\right)
+\lambda K\nabla_N\right]
 \left[ (\rho +p)u_{C}^2\right]\cr
&+&\left[ {1\over 2} -\lambda\nabla^2_C\right]\rho
-{1\over d}\lambda K^2\left(\rho +p\right)\left( u_{C}^2 +d\right)
-{1\over 2}K\sigma
=0~.
\label{eq:nn}
\ea

We treat analogously the equations of the perfect fluid in the bulk.
The equation of motion, Eq.~(\ref{eq:fluid}), can be combined with the
continuity equation, Eq.~(\ref{eq:continuity}), to yield
\ba
\nabla _{\nu}\left[
\left( 1 +\lambda \dim 5 R\right)(\rho +p)
 \left( g_{\mu\nu} +u_{\mu}u_{\nu}\right)\right]
-\left( 1 +\lambda \dim 5 R\right)g_{\mu\nu}\nabla_{\nu}\rho =0.
\label{eq:fluid+continuity}
\ea
Substituting the expression for $\dim 5 R$ in Eq.~(\ref{eq:R^5}), we
integrate both the parallel and the orthogonal components along the
$N$ direction to obtain the corresponding boundary conditions on the
brane.
From the parallel component we find that
\ba
0 &=&\int ^{+\delta}_{-\delta}dN \nabla_{N}\biggl\{
-2\lambda \nabla_{A}\left[
K(\rho +p)\left( g_{AB} +u_{A}u_{B}\right)\right]\cr
&&+\lambda K^2( \rho +p)u_{A}u_{N}\left[
 g_{AB}\left( 1+{1\over d}\right)
 +{ {\lambda(\rho +p)}\over {d\left( M_{(5)}^3 -\lambda \rho\right)}}
   \left( du_{A}u_{B} -g_{AB}u_{C}^2\right)\right]\cr
&&+\left[ 1 +\lambda\left( R -K^2 -K_{CD}K^{CD} -2K_{,N}\right)\right]
 (\rho +p)u_{N}u_{B}
+2\lambda Kg_{AB}\nabla_{A}\rho\biggr\}~.
\ea
Here we encounter a third derivative along the $N$ direction of
the induced metric $g_{AB}.$ For a continuous metric, the first derivate
can be discontinuous, the second derivative can
be a delta-like singularity and consequently the third derivative can
be a double-peaked delta. When we integrate in $N$ we are left
with a term in $K_{,N},$ which is proportional to the second
derivative of the metric and thus potentially singular, evaluated at
the end points along the normal direction which define the thickness of the
brane. Due to the $\mathbb{Z}_{2}$--symmetry, however,
whereas the delta singularity is even about the brane, with
$K(N =-\delta) =-K(N =+\delta)$ and thus $\int dN \nabla_{N}K =2K,$
the double-peaked delta is odd, with
$K_{,N}(N =-\delta) =+K_{,N}(N =+\delta)$ and thus
$\int dN \nabla_{N}\nabla_{N}K =0.$
Consequently, when integrated along $N,$
only odd order derivatives of the metric along $N$ jump across the
brane and thereby relating with the singular matter distribution at
the location 
of the brane, while even order derivatives cancel at the end points.
However, the term in question contains also the factor $u_{A}u_{N}$
which is odd about the brane, thus causing the integral to survive and yield
$\int dN \nabla_{N}\left(K_{,N}u_{A}u_{N}\right) =2K_{,N}u_{A}u_{N}$
at $N=+\delta.$ On the other hand, since the boundary condition in
Eq.~(\ref{eq:israel:nn}) imposes that 
either $u_{A}=0$ or $\rho +p =0$ or $K=0,$ then regardless the case
this term vanishes on the brane.
Then, the parallel component of the boundary condition becomes
\ba
&-&2\lambda \nabla_{A}\left[
K(\rho +p)\left( g_{AB} +u_{A}u_{B}\right)\right]
+2\lambda Kg_{AB}\nabla_{A}\rho \cr
&+&( \rho +p)u_{B}u_{N}
\left[
1 +\lambda\left\{ R
+{1\over d}K^2\left(
 1+{{\lambda(\rho +p)}\over {\aalpha -\lambda \rho}}u_{C}^2(d-1)\right)
-K_{CD}K^{CD}\right\}\right]
=0~.\qquad
\label{eq:israel:a}
\ea
Substituting the boundary condition in Eq.~(\ref{eq:israel:a}) back in the
parallel projection of Eq.~(\ref{eq:fluid+continuity}),
and using also the energy conservation condition in
Eq.~(\ref{eq:conservation}), we find for the induced equation for the
fluid on the brane
\ba
&&\nabla_{A}\left[
\left( 1 +\lambda\left\{ R -2K^2 -K_{CD}K^{CD}
+2K\nabla_{N}\right\}\right)
 (\rho +p)\left( g_{AB} +u_{A}u_{B}\right)\right]\cr
%&+&2\lambda\nabla_{A}\left[
%K\nabla_{N}\left(
% (\rho +p)\left( g_{AB} +u_{A}u_{B}\right)\right)\right]\cr
&+&2\lambda K(\rho +p)\left[
 u_{A}u_{B}\left( K_{AB|C} -K_{AC|B}\right)
 +u_{A}u_{N}\left( K_{CD}K_{DC}g_{AB} +K_{AD}K_{DB}\right)\right] \cr
&-&\lambda K^2\nabla_{A}\left[
(\rho +p)\left( u_{A}u_{B} +g_{AB}\right)\right]\cr
&-&\lambda K^2\left[
g_{AB}\left( 1 +{1\over d}\right)
+{1\over d}{ {\lambda(\rho +p)}\over {\aalpha -\lambda\rho}}
  \left( du_{A}u_{B} -g_{AB}u_{C}^2\right)
\right]
 \nabla_{N}\left[(\rho +p)u_{A}u_{N}\right]\cr
&-&\left[
2\lambda K\left( K_{AB} -g_{AB}K\right)
+g_{AB}\left(1 +\lambda\left\{ R -K^2 -K_{CD}K^{CD}\right\}
\right)\right]\nabla_{A}\rho =0~.
\label{eq:a}
\ea
The orthogonal component yields a trivial matching condition because
of the continuity conditions across the brane of the quantities
involved. We conclude that this component will only be important for the
propagation of the fluid off the brane and across the bulk. The
propagation on the brane is solely described by Eq.~(\ref{eq:a}), which
contains already the continuity condition in Eq.~(\ref{eq:continuity}),
with the possibility of propagation off into the bulk being 
constrained by the conservation condition in Eq.~(\ref{eq:conservation}).

These are the induced equations on the
brane and can be solved the following way. The effective
equations of motion in Eqs.~(\ref{eq:ab}) and (\ref{eq:a}) consist of a
coupled system which must be solved together for the induced metric
and for $\nabla_{N}\left[ (\rho +p)\left(u_{A}u_{B} +g_{AB}\right)\right].$
With these results, we can then solve the stress-energy conservation 
condition in Eq.~(\ref{eq:conservation}),
derived from the $(AN)$ component of the Einstein equation,
and the $(NN)$ component of the Einstein
equation in Eq.~(\ref{eq:nn})
together for $\rho$ and the extrinsic curvature, constrained by the
matching conditions in Eqs.~(\ref{eq:israel:ab}) and
(\ref{eq:israel:a}) which then allow
%Finally, we use Eq.~(\ref{eq:continuity})
to find the functional form of $p$ in terms of $\rho.$

From these equations we observe that the coupling of the curvature to
the matter Lagrangian density yields a contribution to the effective Newton's
constant on the brane. Moreover, it is only through the non-minimal
coupling that matter in the bulk interacts with that 
localized on the brane, in this case the tension $\sigma$ only.
Notice that if $\lambda=0$, i.e.
in the absence of a non-minimal coupling of the curvature to the matter
Lagrangian density, 
Eqs.~(\ref{eq:israel:a}) and (\ref{eq:a}) read respectively
\ba
\rho +p = 0
\label{c.c.}
\ea
and
\ba
\nabla_{A}(\rho +p)\left( g_{AB} +u_{A}u_{B}\right)-g_{AB}\nabla_{A}\rho =0~.
\label{eq.all}
\ea
This means that the presence of a perfect fluid in the bulk space
induces on the  brane an equation of state characteristic of a
cosmological constant,
%implies that $\rho$ and $p$ are related through an
%equation of state of the cosmological constant type
without, however, the quantities $\rho$ and $p$ characterizing the fluid
being necessarily constant.

\section{Results
\label{sec:results}}

In this section we proceed to study the set of equations derived above
for the three particular cases encompassed by the boundary condition derived
from the $(NN)$ component of the Einstein equation,
Eq.~(\ref{eq:israel:nn}).

\subsection{Reflecting Boundary Condition: $u_{A}=0, \nabla_{N}u_{N}=0$}

One of the cases that we can consider is that of the brane
consisting of a reflecting surface for the incoming fluid flux from
the bulk. This translates into setting Dirichlet boundary conditions to
the parallel component of the fluid velocity $u_{A}=0$ and Neumann
boundary conditions to the normal component $\nabla_{N}u_{N}=0$ at the
position of the brane. Moreover, the $\mathbb{Z}_{2}$--symmetry
implies that $\nabla_{N}u_{A}=0,$ whereas boost and translation
invariance on the brane implies that $\nabla_{B}u_{A}=0.$

The equations derived in the previous section become as
follows. The $(AB)$ and the $(NN)$ components of the Einstein equation
are given respectively by
\ba
&&\left( \aalpha -\lambda \rho\right)
 \left[
  G_{AB} -KK_{AB} +{1\over 2}g_{AB}\left( K_{CD}K_{CD} +K^2\right)\right]
+g_{AB}\Lambda _{(5)}\cr
&-&\left[{1\over 2}\left( 1 +\lambda\left\{ R -K^2 -K_{CD}K_{CD}\right\}\right)
+\lambda K\nabla_{N}\right](\rho +p)g_{AB} \cr
&+&\left[ {1\over 2}g_{AB}
 +\lambda\left( \nabla_{A}\nabla_{B} -g_{AB}\nabla_{C}^2\right)\right]\rho
=0~,
\label{eq:ab:caseA}
\\
\cr
&&\left( \aalpha -\lambda \rho\right)
 {1\over 2}\left[ -R +K^2\left( {2\over d} -1\right) -K_{CD}K_{CD}\right]
+\Lambda _{(5)}
\cr
&+&\left[ {1\over 2} -\lambda\nabla_{C}^2\right]\rho
-\lambda K^2\left(\rho +p\right) -{1\over 2}K\sigma =0~;
\label{eq:nn:caseA}
\ea
the parallel component of the induced fluid equation and the
condition of stress-energy conservation on the brane respectively by
\ba
&&\nabla_{A}\left[
\left( 1 +\lambda\left\{ R -2K^2 -K_{CD}K^{CD}
+2K\nabla_{N}\right\}\right)
 (\rho +p)g_{AB}\right]
-\lambda K^2g_{AB}\nabla_{A}(\rho +p)\cr
&-&\left[2\lambda K\left( K_{AB} -g_{AB}K\right)
+g_{AB}\left( 1 +\lambda\left\{ R -K^2 -K_{CD}K^{CD}\right\}
\right)\right]\nabla_{A}\rho =0~,
\label{eq:a:caseA}
\\
\cr
&&{1\over d}(d -1)\left( \aalpha -\lambda \rho\right)\nabla_{B}K
-\lambda\left( K_{AB} -{1\over d}g_{AB}K\right)\nabla_{A}\rho =0~.
\label{eq:conservation:caseA}
\ea
We present also the matching conditions from both the $(AB)$ component
of the Einstein equation, Eq.~(\ref{eq:israel:ab}), and the parallel
component of the fluid equation, Eq.~(\ref{eq:israel:a}), respectively
\ba
&&\left( \aalpha -\lambda \rho\right)\left( -K_{AB} +g_{AB}K\right)
+\lambda K(\rho +p)g_{AB} -\lambda g_{AB}\nabla_{N}\rho
=-{1\over 2}\sigma g_{AB}~,%\\
\label{eq:israel:ab:caseA}
%\cr
\ea
and
\ba
&&(\rho +p)\nabla_{A}K +K\nabla_{A}p=0~.
\label{eq:israel:a:caseA}
\ea 
In this case, the extrinsic curvature is intertwined with both the
energy density and the pressure of the fluid via the non-minimal
coupling $\lambda,$ being in addition sourced by the brane tension
$\sigma.$ As in the case described next, the role of the brane tension
seems superfluous for generating the discontinuity of the bulk
geometry at the position of the brane and thus accounting for the
singular presence of the brane in the bulk space.  Although the system
is intricately entangled, we have just enough equations and constraints
to be able to solve it unambiguously given initial conditions. The
procedure follows that suggested above for the general system. For
further insight into a possible nature of such bulk field, we draw
the attention to the next case.

\subsection{Cosmological Constant Fluid: $\rho +p = 0$}

Another case contemplated by Eq.~(\ref{eq:israel:nn}) is that of the
fluid equation of state induced on the brane being $\rho +p=0,$ which
corresponds to the bulk fluid inducing a cosmological constant on the
brane.
The $(AB)$ and $(NN)$ components of the induced Einstein equations
become respectively
\ba
&&\left( \aalpha -\lambda \rho\right)
 \left[
  G_{AB} -KK_{AB} +{1\over 2}g_{AB}\left( K_{CD}K_{CD} +K^2\right)\right]
+g_{AB}\Lambda _{(5)}\cr
&-&\lambda \left( u_{A}u_{B} +g_{AB}\right)K\nabla_{N}(\rho +p)\cr
%+{1\over 2}g_{AB}\rho
&+&\left[ {1\over 2} g_{AB}
+\lambda\left( \nabla_{A}\nabla_{B}
-g_{AB}\nabla_{C}^2\right)\right]\rho
=0~,
\label{eq:ab:caseB}
\\
\cr
&&\left( \aalpha -\lambda \rho\right)
 {1\over 2}\left[ -R +K^2 \left( {2\over d} -1\right) -K_{CD}K_{CD}\right]
+\Lambda _{(5)}
\cr
&+&\lambda u_{C}^2K\nabla_{N}(\rho +p)
%+{1\over 2}\left(\rho -K\sigma\right)
+\left[ {1\over 2} -\lambda\nabla_{C}^2\right]\rho
-{1\over 2}K\sigma
=0~,
\label{eq:nn:caseB}
\ea
whereas the parallel component of the fluid equation and the
stress-energy conservation condition become respectively
\ba
&&\left( 1 +\lambda\left\{ R -2K^2 -K_{CD}K^{CD}
+2K\nabla_{N}\right\}\right)
\left[ \left( u_{A}u_{B} +g_{AB}\right)\nabla_{A}(\rho +p)\right]\cr
%2\lambda K
% \left( u_{A}u_{B} +g_{AB}\right)
%  \nabla_{N}\nabla_{A}(\rho +p)\cr
&-&\lambda K^2\left(u_{A}u_{B} +g_{AB}\right)\nabla_{A}(\rho +p)\cr
&+&\lambda\left[
2\left(u_{A}u_{B} +g_{AB}\right)\nabla_{A}K
+2K\nabla_{A}\left( u_{A}u_{B}\right)
-K^2\left( 1 +{1\over d}\right)u_{B}u_{N}\right]\nabla_{N}(\rho +p)
\cr
&-&\left[2\lambda K\left( K_{AB} -g_{AB}K\right)
+g_{AB}\left( 1 +\lambda\left\{ R -K^2 -K_{CD}K^{CD}\right\}
\right)\right]\nabla_{A}\rho
=0~,\qquad
\label{eq:a:caseB}
\\
\cr
&&{1\over d}(d -1)\left( \aalpha -\lambda \rho\right)\nabla_{B}K
-\lambda\left( K_{AB} -{1\over d}g_{AB}K\right)\nabla_{A}\rho \cr
&-&\lambda K\left(u_{A}u_{B} -u_{C}^2g_{AB}\right)\nabla_{A}(\rho +p)
=0~.
\label{eq:conservation:caseB}
\ea
The corresponding matching conditions are
\ba
&&\left( \aalpha -\lambda \rho\right)\left( -K_{AB} +g_{AB}K\right)
-\lambda g_{AB}\nabla_{N}\rho
=-{1\over 2}\sigma g_{AB}~
\label{eq:israel:ab:caseB}
\ea
and
\ba
&& \nabla_{A}p
+u_{A}u_{B}\nabla_{B}(\rho +p)
=0~.
\label{eq:israel:a:caseB}
\ea

We now proceed to investigate this case in more detail.
Since $\rho +p=0$ everywhere on the brane, then from boost and translation
invariance we must have that $\nabla_{A}(\rho +p)=0$ everywhere on the
brane. Since both $\rho +p$ and its first derivative along the
parallel directions to the brane vanish on the brane, then so must
vanish its second derivative. However, terms in $\nabla_{N}(\rho +p)$
or $\nabla_{N}\nabla_{A}(\rho+p)$ do not necessarily vanish.
From Eq.~(\ref{eq:israel:a:caseB}) it follows that
$\nabla_{A}p=\nabla_{A}\rho=0,$
and from Eq.~(\ref{eq:conservation:caseB}) that $\nabla_{B}K=0.$
Moreover, since $\nabla ^2(\rho +p)=0$ implies that 
$\nabla ^2\rho =-\nabla ^2p,$ whereas $\nabla \rho =\nabla p$ implies
that $\nabla ^2\rho =\nabla ^2p,$ we must have that 
$\nabla ^2\rho =\nabla ^2p=0.$ 
We can then solve the Einstein, Eq.~(\ref{eq:ab:caseB}), 
and the fluid equation, Eq.~(\ref{eq:a:caseB}),
iteratively for the evolution of the induced metric and
$\nabla_{N}(\rho +p)$ on the brane, with the sole ambiguity residing
in the fluid velocity in the bulk. The value of the extrinsic
curvature will then be given by Eq.~(\ref{eq:nn:caseB}).
We also note that with the bulk field behaving on the brane as an
effective cosmological constant the role of the brane tension becomes
superfluous. 

Furthermore, should we allow for
$\nabla_{A}(\rho +p)\not=0,$ then the discontinuity of the extrinsic
curvature across the brane would be further sourced by the evolution
of $\rho+p$ on the brane and thus generated dynamically according to
the equation of motion for the fluid induced on the brane. 
This would be an interesting idea to pursue further.
Despite the increased complexity of the problem, the coupled system
would still be solvable.

\subsection{Vanishing Extrinsic Curvature: $K=0$}

The remaining case is that of a vanishing extrinsic curvature, where
the $(AB)$ and $(NN)$ components of the effective Einstein equations
become
\ba
&&\left( \aalpha -\lambda \rho\right)G_{AB}
-{1\over 2}\left(1 +\lambda R\right)
 \left( u_{A}u_{B} +g_{AB}\right)(\rho +p)
+g_{AB}\Lambda _{(5)}\cr
&+&\left[ {1\over 2} g_{AB}
+\lambda\left( \nabla_{A}\nabla_{B}
-g_{AB}\nabla_{C}^2\right)\right]\rho
=0~,
\label{eq:ab:caseC}
\\
\cr
&&-\left( \aalpha -\lambda \rho\right)
 {1\over 2}R
+{1\over 2}\left( 1 +\lambda R\right)(\rho +p)u_{C}^2
+\Lambda _{(5)}
+\left[ {1\over 2} -\lambda\nabla_{C}^2\right]\rho
=0~,
\label{eq:nn:caseC}
\ea
and the effective equations for the fluid
\ba
&&
\nabla_{A}\left[ (\rho +p)\left( u_{A}u_{B} +g_{AB}\right)\right]
-g_{AB}\nabla_{A}\rho =0~,
\label{eq:a:caseC}
\\
\cr
&&(\rho +p)\left( u_{A}u_{B} +g_{AB}\right)\nabla_{A}K
=0~,
\label{eq:conservation:caseC}
\ea
where we used the Israel matching condition
\ba
\lambda\nabla_{N}\rho ={1\over 2}\sigma~.
\label{eq:israel:ab:caseC}
\ea
For $K=0,$ the brane tension is
supported by the discontinuity in the energy density of the bulk fluid
across the two sides about the brane.
The matching condition for the fluid equation then yields
\ba
(\rho +p)u_{B}u_{N}\left( 1+\lambda R\right)=0~
\label{eq:israel:a:caseC}
\ea
%which leads to an asymptotic cosmological constant equation of state
%for a universe with a vanishing small scalar curvature. BOO
and consequently $R=-1/\lambda.$ \footnote{The other possibility,
$\rho +p=0,$ would be but a particular case of the bulk fluid behaving
as a cosmological constant on the brane, as discussed previously.}
Then Eq.~(\ref{eq:nn:caseC}) reduces to 
\ba
\nabla_{C}^2\rho 
-{1\over \lambda}\left( 
 {M^3_{P(5)}\over \lambda} +\Lambda _{(5)}\right)=0~.
\label{eq:nn:caseC2}
\ea
This equation can be solved for $\rho$ given initial conditions 
at the intersection of the brane with the bulk
past infinity, obtaining the evolution on the brane of the energy
density of the bulk perfect fluid in terms of the parameters of the
bulk space. The solution must also be subject to the reproduction of 
a consistent bulk cosmological constant, in the case of anti-de Sitter 
$\Lambda _{(5)}<0.$ 
Upon substitution of Eq.~(\ref{eq:nn:caseC}), 
Eq.~(\ref{eq:ab:caseC}) becomes
\ba
\left( M^3_{P(5)} -\lambda\rho\right)\left(
G_{AB} -{1\over {2\lambda}}g_{AB}\right)
+\lambda\nabla_{A}\nabla_{B}\rho
=0~.
\label{eq:ab:caseC2}
\ea
From Eq.~(\ref{eq:conservation:caseC}), we must have that
$\nabla_{A}K=0.$
We can then use the solution for $\rho$ to solve
Eq.~(\ref{eq:ab:caseC2}) for $g_{AB}$ and
Eq.~(\ref{eq:a:caseC}) for $p$ given $u_{A}.$
Hence, for $K=0$ the system decouples and can be solved straightforwardly.
This case is, however, too limited since it does not constrain the evolution
of $\nabla_{N}\rho$ capable of generating the brane tension $\sigma$
in Eq.~(\ref{eq:israel:ab:caseC}).

\section{Conclusions\label{sec:conclusions}}

In this paper we have considered a modified gravity model where the
curvature scalar couples non-minimally to the matter Lagrangian
density, which here we realize for the case of a perfect fluid. As
discussed in Ref. \cite{Bertolami:2007gv}, in four dimensions this
model can potentially  account for the observed rotation curves of
galaxies without recourse to dark matter and suggests a solution to
the Pioneer anomaly. In addition to this coupling, we have also
considered an extra spatial dimension, so that our spacetime is
embedded in a five-dimensional bulk space where gravity is described by
a five-dimensional Einstein equation.

We find that the resulting model is well defined for the
considered physical variables and can be solved for given initial
conditions.
The new terms that arise from the bulk-brane decomposition yield quite
interesting consequences. We found three particular cases which
conform to the matching conditions upon the
assumption of $\mathbb Z_2$--symmetry about the position of the
brane and investigated their contribution to the intricately
entangled system of equations of motion therein induced. These
cases are the following:
a) reflecting boundary conditions on the brane,
$u_{A}=\nabla_{N}u_{N}=0;$
b) an induced fluid equation of state characteristic of a cosmological
constant, $\rho +p=0;$ and
c) a vanishing extrinsic curvature, $K=0.$

Both cases a) and b)
seem to render the presence of a brane tension $\sigma$ superfluous for
generating the discontinuity of the bulk geometry at the position of
the brane, where the energy density $\rho$ and the pressure $p$ of the
fluid can source the discontinuity of the extrinsic curvature in a
dynamical manner.

Furthermore, case b) can be regarded as a generalization of a cosmological
constant scenario, where the bulk fluid induces on the brane a cosmological
constant capable of supporting the presence of the brane.
This implies that the evolution on the brane of the energy density of the
bulk perfect fluid determines the behaviour of the
cosmological constant term on the brane.
It is well known that an evolution in terms of cosmic time as
$\rho \propto t^{-2}$ is consistent with the value of the vacuum
energy density at present \cite{Bertolami86,Sahni}. Furthermore, this
positive contribution for the brane cosmological constant can have
implications for inflation and for the late time acceleration of the
universe. This contribution can also have a bearing on the
cosmological constant problem since the natural background for
fundamental theories, such as supergravity and superstring/M-theory,
is the anti-de Sitter space, which on the brane requires a
compensating de Sitter contribution. The feasibility of a scenario
along these lines will be considered elsewhere.

Finally, case c) allows for the decoupling of the system of
equations, with the energy density relating with the bulk cosmological
constant $\Lambda_{(5)}.$ The presence of the brane is supported by
the interaction of the brane tension with the discontinuity across the
brane of the fluid energy density, which on the brane is governed
dynamically by a completely defined equation for $\rho.$

\appendix

\section{Tensor Decomposition in Gaussian normal coordinates
\label{sec:decomposition}}

Here we derive the projection of the Einstein tensor along the
parallel and normal directions to the brane, namely
$G_{AB}$, $G_{AN}$ and $G_{NN},$
using the Gaussian normal prescription \cite{Bucher:2004we}
\ba
\dim 5 \grad A \versor B
&=&\Gamma_{~BA}^C \versor C -K_{AB} \versor N ~,\\
\dim 5 \grad N \versor A  &=&K_A^C \versor C ~,\\
\dim 5 \grad A \versor N &=&K_A^C \versor C~.
\ea
The Gaussian coordinates split the bulk space into a four-dimensional
spacetime times a one-dimensional space, so that
\ba
\dim 5 \bf g = \left(\begin{array}{ccc|c}
 & & & \\
 & \bf \dim 4 g & & \\
& & & \\\hline
& & & 1 \\
\end{array}\right) ~.
\ea
Here, $A,$ $B,...$ denote directions parallel to the brane, while $N$
denotes the orthogonal direction.

We aim to obtain decompositions of the form
\ba
\dim 5 G_{AB} =\text{terms without extrinsic curvature}
+\text{terms with extrinsic curvature} ~,
\ea
so that the term without extrinsic curvature will be identified
with $\dim 4 G_{AB}.$ We use the  Einstein equations in the bulk to
eliminate $\dim 5 G_{AB}$ and thus obtain the Einstein equation
parallel to the brane. However, this equation will still contain
terms in the extrinsic curvature, which can be related to other
quantities on the brane through the Israel matching conditions.
In this calculation we use geodesic coordinates on the brane, i.e., on
the metric $\dim 4 g_{AB}$. However, this does not mean that the
connection components $\Gamma^C_{~AB}$ can be ignored, since their
derivatives are non-vanishing.

The Riemann tensor can be written as
\ba
R(Y,W)Z =\nabla_Y \nabla_W Z -\nabla_W \nabla_Y Z
-\nabla_{\left[Y,W\right]}Z ~,
\ea
where $Y,$ $W$ and $Z$ are vector fields, so that in components:
\ba
R^{\mu}_{~\alpha\gamma\beta}\versor {\mu} 
=\left( \grad {\gamma} \grad {\beta}
-\grad {\beta} \grad {\gamma} \right) \versor {\alpha} ~.
\ea
On its turn, the Ricci tensor is given by
\begin{equation}
 R_{\alpha\beta} = dx^{\gamma} (R^{\mu}_{~\alpha\gamma\beta}\versor {\mu}) ~.
\end{equation}

Let us now compute the decomposition of $R_{AB}$. First, one computes
$\dim 5 {R^J}_{ACB}\versor J$~ and  $\dim 5 R^J_{~ANB}\versor J.$~ 
Noticing that the indices $J$ and $L$ refer to all five dimensions, it
follows that
\ba
\dim 5 {R^J}_{ACB}\versor J
&=& \dim 5 \grad C \dim 5 \grad B \versor A -\left( C\leftrightarrow B\right)
\cr
&=& \dim 5 \grad C \left( \Gamma^D_{AB}\versor D -K_{AB}\versor N\right)
-( C\leftrightarrow B)\cr
&\doteq& \Gamma^D_{AB,C}\versor D -K_{AB,C}\versor N
-K_{AB}K_C^D\versor D -\left( C\leftrightarrow B\right) ~.
\ea
The symbol $\doteq$ denotes equality only for the case of geodesic
coordinates on the brane. We also note that
\ba
\dim 4 \grad C ( K_{AB}dx^A \otimes dx^B)
\doteq \partial_C \left( K_{AB}\right) dx^A \otimes dx^B,
\ea
which is equivalent to $K_{AB|C} \doteq K_{AB,C},$
where $|C$ denotes $ \dim 4 \grad C $. Then,
\ba
\dim 5 {R^J}_{ACB}\versor J
&\doteq& \left( \Gamma^D_{AB,C} -\Gamma^D_{AC,B}\right)\versor D 
+\left( K_{AC}K_B^D -K_{AB}K_C^D\right)\versor D \cr
&+&\left( K_{AC|B} -K_{AB|C}\right)\versor N 
~.
\ea
Computing now $\dim 5 R^J_{~ANB}\versor J$ ~we find that
\ba
\dim 5 R^J_{~ANB}\versor J
&=&\dim 5 \grad N \left( \Gamma^D_{~AB}\versor D -K_{AB}\versor N\right)
-\dim 5 \grad B \dim 5 \grad N \versor A \cr
&\doteq& \Gamma^D_{~AB,N} \versor D -K_{AB,N}\versor N
-K_{A|B}^D\versor D +K_A^D K_{DB}\versor N ~.
\ea
The five-dimensional Ricci tensor components can now be written as
\ba
\label{eq:rab}
\dim 5 R_{AB}
&=& dx^L \left( \dim 5 {R^J}_{ALB}\versor J \right)
=dx^C \left( \dim 5 {R^J}_{ACB}\versor J \right)
+dx^N \left( \dim 5 {R^J}_{ANB}\versor J \right)\cr
&\doteq& \Gamma^C_{AB,C} -\Gamma^C_{AC,B} +K_{AC}K_{B}^C -KK_{AB}
-K_{AB,N} +K_A^C K_{CB}\cr
&=&\dim 4 R_{AB} -KK_{AB} +2K_{AC}K^C_B -K_{AB,N} ~,
\ea
where $K = K_{AB}\dim 4 g^{AB}.$
%Although not explicitly, this is now written in a tensorial way,
%meaning that if this is true in geodesic coordinates, it is true
%for any coordinate system. Some steps below, one needs to compute
%$\dim 5 R_A^B$. However, it is not correct to put the index up in
%all the terms since $\partial_N g^{BC} \neq 0$.
Moreover, we compute
\ba
\dim 5 \grad N \left( K_{AB}dx^A dx^B\right)
&=& K_{AB,N}dx^A dx^B
+K_{AB}dx^A \dim 5 \grad N dx^B +K_{AB}dx^B \dim 5 \grad N dx^A\cr
&=& K_{AB,N}dx^A dx^B -2K_{AC}K^C_B dx^A dx^B ~,
\ea
which yields
\ba
K_{AB;N} =K_{AB,N} -2K_{AC}K^{C}_B ~.
\ea
Finally, $\dim 5 R_{AB}$ can be rewritten as
\ba
\dim 5 R_{AB} =\dim 4 R_{AB} -KK_{AB} -K_{AB;N} ~.
\ea
Analogously, the other components of the Ricci tensor are given by
\ba
\dim 5 R_{NN} &=& -K_{CD}K^{CD} -K_{,N} ~,
\label{eq:rnn}\\
\dim 5 R_{NA} &=& K_{A|B}^B -K_{|A} ~,
\ea
so that the curvature scalar is expressed as
\ba
\dim 5 R
=g^{AB}\dim 5 R_{AB} +\dim 5 R_{NN}
=\dim 4 R -K^2 -K_{CD}K^{CD} -2 K_{,N} ~.
\label{eq:R^5}
\ea

Then, the decomposition of the five-dimensional Einstein tensor
results as follows 
\ba
\dim 5 G_{AB}
&=& \dim 5 R_{AB} -{1\over 2} \dim 5 g_{AB} \dim 5 R \cr
&=& \dim 4 G_{AB} +2K_{AC}K^C_B -KK_{AB} -K_{AB,N}\cr
&&+{1 \over 2}\dim 4 g_{AB}\left( K_{CD}K^{CD} +K^2 +2K_{,N}\right) ~,
\label{eq:gc1}\\
\dim 5 G_{NA}
&=& \dim 5 R_{NA} -{1\over 2} \dim 5 g_{NA} \dim 5 R \cr
&=& K_{A|B}^B -K_{|A} ~,
\label{eq:gc2}\\
\dim 5 G_{NN}
&=& \dim 5 R_{NN} -{1\over 2} \dim 5 g_{NN} \dim 5 R \cr
&=& -{1\over 2}\dim 4 R +{1\over 2} \left( K^2 -K_{CD}K^{CD}\right) ~,
\label{eq:gc3}
\ea
Equations (\ref{eq:gc1} -- \ref{eq:gc3}) are known as the Gauss-Codacci
relations.

In order to derive the matching conditions in our model, we point out that
the Einstein equation, Eq. (\ref{eq:Einstein5dim}), contains second
derivatives of the energy density $\rho$. We need to express them in
Gaussian coordinates. Under a transformation of coordinates
$\versor J = \Lambda^\nu_J \versor \nu,~$ where $\Lambda^\nu_J$ are the
components of the versor $\versor J~$ expressed in the $\versor \nu~$ basis,
\ba
\nabla_\mu \nabla_\nu \rho \quad \longrightarrow
&\quad &\Lambda^\mu_I \Lambda^\nu_J \nabla_\mu \nabla_\nu \rho \cr
&=& \Lambda^\mu_I \nabla_\mu \left( \Lambda^\nu_J \nabla_\nu \rho\right)
-\Lambda^\mu_I \left( \nabla_\mu \Lambda^\nu_J \right) \nabla_\nu \rho \cr
&=&\nabla_I \nabla_J \rho - \Gamma^L_{~J I} \nabla_L \rho ~.
\ea
Then, terms of the form $\nabla_\mu \nabla_\nu \rho$ decompose along
the directions parallel and orthogonal to the brane as follows
\ba
(AB) &\qquad & \nabla_A \nabla_B \rho + K_{AB} \nabla_N \rho ~,\\
(AN) &\qquad & \nabla_A \nabla_N \rho - K_A^B \nabla_B \rho ~,\\
(NN) &\qquad & \nabla_N \nabla_N \rho ~.
\ea
Consequently,
\ba
\dim 5 \square \rho = \square \rho +\nabla_N \nabla_N \rho +K \nabla_N \rho ~.
\ea

\vfill
%\newpage

%%%%%%%%%%%%%%%%%%%%%%%%%%%%%%%%%%%%%%%%%%%%%%%%%%%%%%%%%%%%%%%%%%%%%%%%%
\centerline{\bf {Acknowledgments}}

\vspace{0.3cm}

\noindent 
C. C. and J.N.L. thank Funda\c c\~ao para a Ci\^encia e a Tecnologia -
FCT (Portuguese Agency) for financial support, 
under the fellowships /BPD/18236/2004 and SFRH/BD/36290/2007 respectively.
O. B. acknowledges the partial support of the FCT project
POCI/FIS/56093/2004.

%\vfill

%%%%%%%%%%%%%%%%%%%%%%%%%%%%%%%%%%%%%%%%%%%%%%%%%%%%%%%%%%%%%%%%%%%%%
%\newpage

%\bibliographystyle{JHEP.bst}
%\bibliography{../Thesis}

%\providecommand{\href}[2]{#2}\begingroup\raggedright\begin{thebibliography}{1}

%\endgroup

\end{document}